\documentclass{an}
\usepackage{graphicx}
\usepackage{times}
\Volume{00}              
\Year{0000}              
\Month{00}               
\Pagespan{000}{000}      

\newcommand{\lnls}{log\,N($>$S)-log\,S}
\newcommand{\nh}{N$_{\rm H}$}
\begin{document}
\lhead[\thepage]{A.N. Author: Title}
\rhead[Astron. Nachr./AN~{\bf XXX} (200X) X]{\thepage}
\headnote{Astron. Nachr./AN {\bf 32X} (200X) X, XXX--XXX}

\title{The source content of low galactic latitude XMM-Newton surveys}

\author{C. Motch, O. Herent and P. Guillout \fnmsep\thanks{On behalf of the XMM-Newton Survey Science
Center}}

\institute{UMR 7550 du CNRS, Observatoire Astronomique de Strasbourg, 11 rue de
l'Universit\'e,France}

\date{Received {date will be inserted by the editor}; 
accepted {date will be inserted by the editor}} 

\abstract{ We present results from a project conducted by the Survey Science Center of the
XMM-Newton satellite and aiming at the identification and characterisation of serendipitous EPIC
sources at low galactic latitudes. Deep multi-colour optical imaging and spectroscopic
observations have been obtained in the framework of several observing campaigns carried out at
ING, CFHT and ESO. These observations have lead to a number of optical identifications, mostly
with active stars. We describe the identified source content at low galactic latitudes and
compare stellar populations properties at low and high galactic latitudes with those expected
from stellar X-ray count models.}

\correspondence{motch@astro.u-strasbg.fr}

\maketitle

\section{Introduction}

Thanks to their excellent sensitivity, good image quality over a wide field of view (30\arcmin )
and wide energy range (0.2-12 keV) the EPIC cameras on board XMM-Newton
allow X-ray surveys with an unprecedented combination of large area and depth. This
potential was recognised by ESA in setting up a dedicated XMM-Newton Survey Science Center
(SSC) to facilitate the exploitation of the XMM-Newton Serendipitous Sky Survey by
providing a public archive of data products and carrying out a carefully coordinated
follow-up programme to characterise the overall X-ray source population (Watson et al.
2001). 

In this paper we present results from an ongoing SSC optical campaign whose
aim is to provide a reference frame for the characterisation of all
low galactic latitude EPIC serendipitous detections. Beyond proper identification and
classification, our objective is to study source properties, quantify their relative
populations and contributions to the overall soft and hard X-ray galactic emission.
XMM-Newton measurements are not strongly biased by interstellar extinction, as were
earlier studies based on ROSAT observations restricted to the softer 0.5-2 keV band.
Although similar in X-ray flux to the deepest ROSAT pointings, the galactic landscape
unveiled by XMM-Newton differs from that seen by ROSAT. Opening new parameter spaces
the present survey offers the possibility to reveal completely new classes of sources.

Many individual source types are known to contribute to the galactic X-ray source
demography in the medium to low X-ray luminosity regime ($10^{28-36} 
\rm~erg~s^{-1}$). These include, late and early-type stars, cataclysmic variables
(CVs), RS~CVns and various species of X-ray binaries. Other kinds of rare emitters
such as isolated neutron stars or low-luminosity but long-lived evolutionary states of
classical XRBs may be found.

\section{Observational material}   

Target fields (see Table \ref{fields}) were selected for their good X-ray quality (depth and
low background) and chosen in directions void of extended diffuse emission or bright sources.
We also avoided atypical regions such as known star forming complexes. Mainly based on EPIC
pn data, source detections were visually screened and positions were in most cases corrected
for residual attitude errors by cross-correlating them with large optical catalogues. In
order not to exclude extreme sources we retained detections in all energy bands. 

Optical data were collected in the framework of the AXIS project (Barcons et al. 2002)
performed at the Observatorio del Roque de los Muchachos in the Canary Islands and from
programmes conducted at CFHT and at ESO. The wide field camera at INT, the CFH12K at CFHT and
the WFI at ESO-MPG 2.2m provided wide-field imaging in the g', r', i', z, R, I and H$\alpha$ /
H$\alpha$ continuum from which high spatial resolution mapping of the error circles and
candidate priorities for spectroscopic observations were derived. Medium to low resolution
spectroscopy was obtained on 4m-class telescopes using multi-fiber instruments (WHT/WYFFOS)
or long slit spectrographs (WHT/ISIS, ESO 3.6m/EFOSC2). 

\begin{table}[h]
\caption{Galactic target fields}
\begin{tabular}{cccccc}\hline
Field & RA & Dec & l & b& Gal \nh \\
\hline
WR 110 & 18h07 & -19\degr 23\arcmin & 10.8\degr & +0.4\degr & 5.5 10$^{22}$ \\
Ridge3 & 18h27 & -11\degr 29\arcmin & 20.3\degr & +0.0\degr & 1.1 10$^{23}$ \\
G21.5-09 & 18h33 & -10\degr 34\arcmin & 21.5\degr & -0.9\degr & 9.0 10$^{22}$ \\
Z And & 23h33 & +48\degr 49\arcmin & 109\degr & -12.1\degr & 1.4 10$^{21}$ \\
\hline
\label{fields} 
\end{tabular}
\end{table}

The total area surveyed so far is $\sim$ 0.8\,deg$^{2}$. With exposure times in the
range of 10 to 30 ksec we reach flux limits of $\sim$ 4 10 $^{-15}$ erg cm$^{-2}$
s$^{-1}$ in the 0.5 to 2.0 keV soft band and of $\sim$ 3 10 $^{-14}$ erg cm$^{-2}$
s$^{-1}$ in the 2 to 10 keV hard energy range. The present survey can thus be
qualified as medium sensitivity in being about 10 times shallower than the Chandra deep
galactic survey of Ebisawa et al. (2002) but 10 times deeper than the medium ROSAT
survey of Morley et al. (2001) or than the faint ASCA galactic plane survey of
Sugizaki et al. (2001). 

In order to compare stellar population properties in the galactic plane with those observed at
high galactic latitudes we also use stellar identifications obtained in the course of the high
$|b|$  AXIS programme (Barcons et al. 2002). The sample consists of 34 stars extracted from 18
XMM-Newton fields ($|b|$ = 23\degr $-$ 80\degr , average value 50\degr ) in a total survey
area of 2.3 deg$^{2}$. The flux limit is 2 10 $^{-14}$ erg cm$^{-2}$ s$^{-1}$ in the 0.5 to 4.5
keV band slightly above that used in the galactic plane.

\section{Identification Strategy}

Identification of XMM-Newton sources in the Milky Way can be made difficult by optical
crowding. Matching the pattern of EPIC sources with the USNO A-2 catalogue and then
correcting for residual attitude errors can shrink the 90\% confidence level error radius
down to $\sim$ 2.5\arcsec . However if field stellar density is too high, no such
adjustment can be made and the 90\% confidence level error radius remains of the order of
4.5\arcsec . Based on our wide field imaging we estimate that at $b$ $\sim$ 0\degr \ the
probability to find at random a star brighter than R $\sim$ 16 - 17 in a typical EPIC error
circle is low enough (5\%) that we can accept the identification on the basis of positional
coincidence. However, in many cases this argument cannot be used and we have to collect
optical spectra of typically 3-4 candidates brighter than R $\sim$ 22 before possibly finding a
reliable identification. At the sensitivity of our survey, stellar counterparts fainter than
R $\sim$ 16 - 17 are mostly late K-M stars. Balmer emission is found in all active M stars
and in a fraction of the most active K stars. Cataclysmic variables and Be/X-ray binaries are
also H$\alpha$ emitters. Our narrow band H$\alpha$ and H$\alpha$ continuum imaging together with
broad band colour information can thus be used to efficiently prioritize optical candidates
for spectroscopy (Herent et al. 2002). We also discarded sources with hardness ratios
consistent with those expected from an absorbed background AGN and with only faint optical
candidates in the error circle. This selection is efficiently done in directions of high
absorption and still leaves a large range of possible \nh \ (or intrinsic spectral hardness)
for galactic objects.

\begin{table*}[ht]
\caption{Statistics of optical identifications at low galactic latitudes}
\begin{tabular}{lccccc}\hline
Field & G21.5-09 & Ridge3 &  WR 110 & Z And & Total\\ \hline
$b$   & -0.9\degr & 0.0\degr & +0.4\degr & -12.1\degr & \\
Area & 0.27 & 0.18 & 0.18 & 0.18 & 0.81 \\
Number of sources & 77 & 30 & 58 & 38 & 203 \\
Stellar coronae & 15 (19\%) & 14 (47\%) & 29 (50\%) & 9 (24\%) & 67 \\
Accreting candidates & 2 & 0 & 1 & 0 & 3 \\
Extragalactic sources & 1 & 0 & 0 & 2 & 3 \\
Unidentified & 59 (77\%) & 16 (53\%) & 28 (48\%) & 27 (71\%) & 130 \\
\hline
\label{ids} 
\end{tabular}
\end{table*}

\section{Optical identifications}

We list in Table \ref{ids} the statistics of optical identifications in each field and for the
entire galactic sample. The limiting magnitude defined here as the faintest magnitude at
which we can detect emission lines typical for a CV or a Me star depends on the instrumental
setting and is R $\sim$ 19 for the G21.5-09 and Z And fields and R $\sim$ 21 for the Ridge 3 and
WR 110 fields. The vast majority of optical identifications are with active coronae. Because of
the relatively high galactic latitude and correspondingly low \nh \ it was still possible to
spectroscopically identify two AGN in the Z And field. At lower latitudes, galactic absorption
becomes too large for our instrumentation. The only low $b$ extragalactic source is in the
field of G21.5-09 (XMMU J183225.4-103645, Nevalainen et al. 2001) and was identified as an X-ray
bright cluster of galaxies seen through over 50 magnitudes of absorption in the visual on the
basis of X-ray source extent and energy distribution.

\begin{figure}
\resizebox{\hsize}{!}
{\includegraphics[angle=-90]{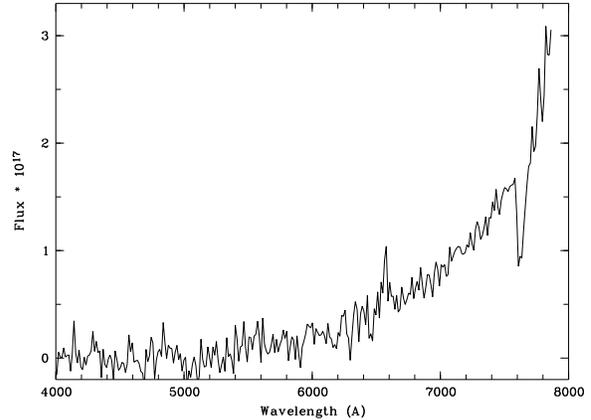}}
\caption{Optical spectrum of a faint accreting candidate in the field of WR 110}
\label{accreting}
\end{figure}

\subsection{Accreting candidates}

In three cases, we detect X-ray emission from a Be star at a level which is slightly above that
expected for 'normal' early type stars (Cassinelli et al. 1994). In addition to the case of SS
397 already mentioned in Motch (2000), we find excess X-ray emission from USNO 0750-13549725 
which is the brightest star in the open cluster NGC 6649. At a distance of 1.6 kpc and with
E(B-V) = 1.38 (Turner 1981) the  X-ray luminosity is $\sim$ 4 10$^{32}$ erg s$^{-1}$ (0.5 - 7.5
keV). Finally, we detect H$\alpha$ emission from a faint red star (R = 22.0, R-I=2.6) in the
field of WR 110. The absence of TiO features and the colour index indicate high interstellar
absorption (see Fig. \ref{accreting}). If the counterpart is an intrinsically blue object then
Av is of the order of 12.5 or \nh \ = 2.2 10$^{22}$ cm$^{-2}$, i.e. about half of the total
expected galactic absorption in that direction. The high interstellar reddening and
accordingly large distance rules out a CV nature. A Be star identification seems likely
and the star would then be at $\sim$ 12 kpc implying an X-ray luminosity of 1.3 10$^{33}$ erg
s$^{-1}$ (0.5 - 7.5 keV). Such modest X-ray luminosity excesses may be in part due to errors on
the reddening. We note however, that the numerous Be + WD systems predicted by binary
evolution theory could well be emitting in this range of luminosities (see Raguzova 2001 and
references therein). 

\subsection{The stellar population}

At very low galactic latitudes, the fraction of sources identified with active coronae varies
very significantly with pointing direction. Whereas in the WR 110 and Ridge 3 fields stellar
coronae account for about half of the X-ray sources, the fraction is only 19\% in the field of
G21.5-09 in spite of the fact that source density is similar to that of WR 110. Such a variance
is not unexpected and could reflect the presence of a higher 'local' absorption in the field of
G21.5-09 preventing spectroscopic identification of the globally optically dimmer stellar
population. 

Spectral types, distances and luminosities were derived for most stellar identifications in the
WR 110 and Ridge 3 fields as well as for the high galactic latitude sample (see Table
\ref{sptypes}). 

In both $b$ $\sim$ 0\degr \ fields, the mean B-V excess of X-ray active stars is about 0.6 (or \nh
\ $\sim$ 3.5 10$^{21}$ cm$^{-2}$). We detect active coronae up to 500 pc in Ridge 3 and 1 kpc in
WR 110.  

At the flux level considered here ($\sim$ 4 10 $^{-15}$ erg cm$^{-2}$ s$^{-1}$ in the 0.5 to
2.0 keV band), the low $|b|$ active star population displays a  distribution in spectral types
which is not statistically different from that seen at high $|b|$. The low latitude ROSAT
survey (Motch et al. 1997) also exhibits a similar spectral type distribution in spite of a
factor 25 lower sensitivity (see Table \ref{sptypes}). At low and high latitudes, the observed
distributions match well those expected from X-ray count models (Guillout et al. 1996,
Guillout \& Motch 2002). In particular, we are not yet in a flux regime faint enough for M
stars to dominate at high latitudes. Deeper surveys could however reveal this effect. 

\begin{figure*}[ht]
 {\includegraphics[clip=true,bbllx=40,bblly=453,bburx=540,bbury=710,width=17.5cm]{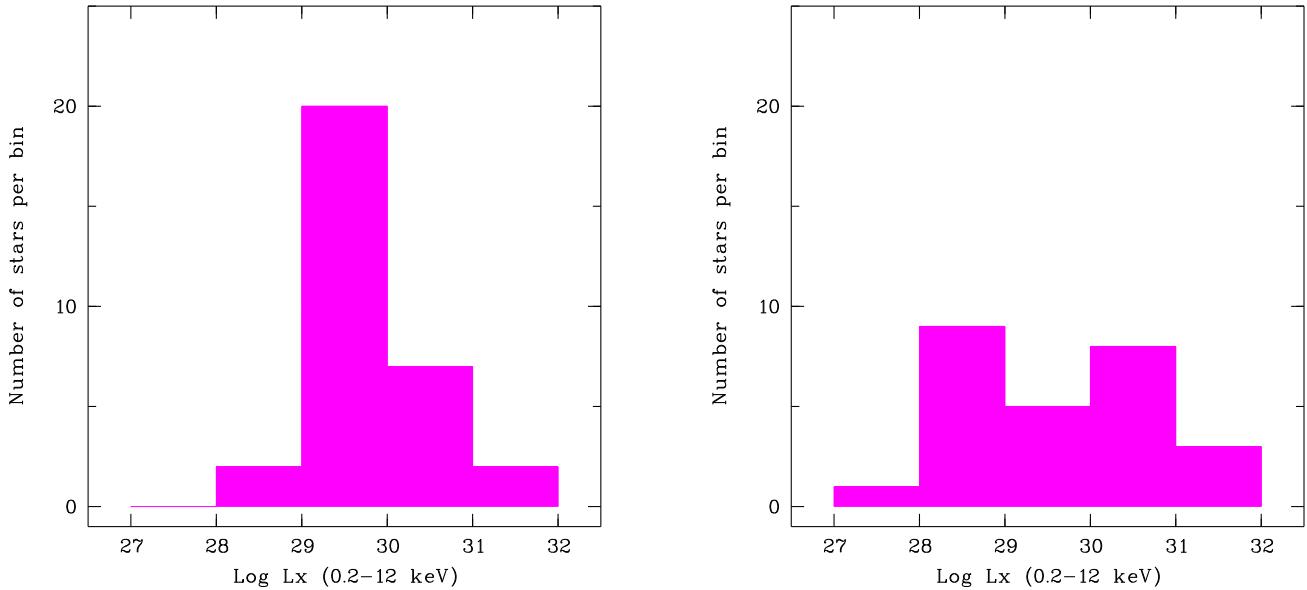}}
\caption{Distribution in X-ray luminosity of active coronae. Left: $|b|$ $\sim$ 0\degr . Right:$\bar{|b|}$
$\sim$ 50\degr }
\label{lx}
\end{figure*}

The distribution in X-ray luminosity of the two $b$ $\sim$ 0\degr \ stellar samples and of the
$\bar{|b|}$ $\sim$ 50\degr \ sample is shown in Fig. \ref{lx}. In the galactic plane, Lx peaks
between 10$^{29}$ and 10$^{30}$ erg s$^{-1}$, typical for stars with ages of the order of that of
the Pleiades ($\leq$ 10$^{8}$ yr, see e.g. Micela et al. 1990). In contrast, the Lx
distribution at high $|b|$ shows two peaks, one between 10$^{28}$ and 10$^{29}$ erg s$^{-1}$ and
one at very high Lx between 10$^{30}$ and 10$^{31}$ erg s$^{-1}$. The low luminosity component
has a solar like X-ray luminosity typical for the relatively old stellar population 
preferentially detected at high latitudes. The origin of the high luminosity component is not yet
clear. It could be due to close binaries such as RS CVns for which synchronized rotation maintains
a strong X-ray luminosity over a long time. Alternatively, a fraction of these high Lx stars
could be misidentifications. 

\begin{table}[hb]
\caption{Distribution in spectral types of active coronae (percents)}
\begin{tabular}{lcccc}\hline

	   \hline
\multicolumn{5}{c}{XMM-Newton : Observed and Model prediction}\\ \hline 
                     & M              &  K             &  G              &  F+A \\
$|b|$ $\sim$ 0\degr  & 28.6 $\pm$ 6.4 & 32.7 $\pm$ 6.7 & 22.4 $\pm$  6.0 &  16.3 $\pm$ 5.3  \\
                     & 24             & 25             & 31              &  20 \\
$|b|$ $\sim$ 50\degr & 29.4 $\pm$ 7.8 & 17.6 $\pm$ 6.5 & 23.5 $\pm$  7.3 &  29.4 $\pm$ 7.8  \\
                     & 27             & 19             & 29              &  25 \\  
\hline
\multicolumn{5}{c}{ROSAT all-sky survey}\\ \hline
                     & M              &  K &  G &  F+A \\
$|b|$ $\sim$ 0\degr  & 19.0 $\pm$ 6.0 & 23.8 $\pm$ 6.5 & 26.2 $\pm$  6.8 &  30.5 $\pm$ 7.1  \\	   
\hline
\label{sptypes} 
\end{tabular}
\end{table}

\begin{figure}
\resizebox{\hsize}{!}
{\includegraphics[bbllx=40,bblly=40,bburx=640,bbury=640,width=8cm,angle=-90]{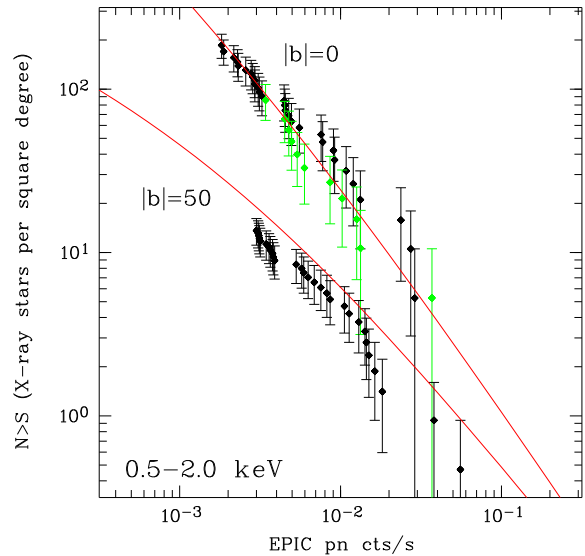}}
\caption{X-ray stellar \lnls \ curves at low and high galactic latitudes. At $|b|$ = 0\degr \ 
the green (or light) plot represents the \lnls \ curve in Ridge 3 and the black one that in WR 110.
Continuous (red) lines show model predictions for $|b|$ = 0\degr and $|b|$ = 50\degr}
\label{starlnls}
\end{figure}

We also show in Fig. \ref{starlnls}  the \lnls \ curve for stars at low and high galactic
latitudes. At EPIC pn count rates of a few 10$^{-3}$ cts/s (0.5 - 2.0 keV), the density of active
stars is about 10 times larger in the galactic plane (here only Ridge 3 and WR 110 fields) than
at high $b$. Predictions of the stellar population models for $b$ = 0\degr \ and 50\degr \ fit
rather well the observed curves. One should however keep in mind the large density variance
observed at low latitudes. As argued above, part of this scatter could be due to complex
absorption structures which are not yet taken into account in the X-ray count model. X-ray
luminous old binaries may also account for a sizeable fraction of the population. Clearly more
stellar identifications are needed in order to build really representative samples.  

\section{Conclusions}

Not unexpectedly, our observations confirm the well known concentration of young active stars  in
the galactic plane. Because of the steep decline of coronal activity with age, X-rays highlight
the young stellar population. Whereas at optical wavelength stars younger than 1\,Gyr make up
only $\sim$ 10\% of the total stellar population and do not exhibit marked signatures, they
account for about half of the X-ray detections, the exact fraction depending on galactic latitude
and limiting flux. The large sensitivity of XMM-Newton allows to sample young stellar populations
up to 1 kpc or more as demonstrated in the WR110 field. At these distances, a number of effects
expected from X-ray stellar populations models become readily visible and can be used to put very
significant constraints on the dynamical properties of young stars, especially the evolution of
scale height with age, and on the SFR during the last $\sim$ 2 Gyr. Among these effects are the
sharp decrease of number count with increasing galactic latitudes, the rising fraction of old
stars at high $b$ and the increasing fraction of M stars at faint fluxes.

\end{document}